# Magnetic domain wall based synaptic and activation function generator for neuromorphic accelerators

Saima A Siddiqui[1], Sumit Dutta[1], Astera Tang[2], Luqiao Liu[1], Caroline A Ross[2], Marc A Baldo[1]

[1]Department of Electrical Engineering and Computer Science, Massachusetts Institute of Technology, Cambridge, MA

[2]Department of Materials Science and Engineering, Massachusetts Institute of Technology, Cambridge, MA

**Magnetic domain walls are information tokens in both logic and memory devices, and hold particular interest in applications such as neuromorphic accelerators that combine logic in memory. Here, we show that devices based on the electrical manipulation of magnetic domain walls are capable of implementing linear, as well as programmable nonlinear, functions. Unlike other approaches, domain-wall-based devices are ideal for application to both synaptic weight generators and thresholding in deep neural networks. Prototype micrometer-size devices operate with 8 ns current pulses and the energy consumption required for weight modulation is ≤ 16 pJ. Both speed and energy consumption compare favorably to other synaptic nonvolatile devices, with the expected energy dissipation for scaled 20-nm-devices close to that of biological neurons.**

Deep neural networks[1,2] mimic the synaptic and activation functionality of human neurons using repeated applications of linear filters interspersed by nonlinear decision functions. Among other applications, deep neural networks offer promising solutions to image[3], speech[4] and video[5] recognition. Training is performed using a backpropagation learning procedure[6], with the filter weights updated continuously like synapses, until the calculated output matches the desired output.

In between the filters, the backpropagation algorithm relies on the finite derivative of the nonlinear decision function to calculate the error.

Hardware implementations of deep neural networks using non-von Neumann architecture offer significant performance advantages over software-based approaches using traditional von Neumann architecture[7]. Depending on the problem and accuracy requirement, a network may require hundreds of layers and each of the layers may consist of thousands of connections[8]. Shuttling data in a von Neumann architecture back and forth from memory to a processor is energy-intensive. Rather, it may be preferable to implement the computation using nonvolatile elements embedded within memory units, an approach known as in-memory computing[9]. Nonvolatile memory is especially well-suited to energy efficient storage of weight information and caching backpropagation training values.

Various nonvolatile devices have been proposed to implement the linear operations in a deep neural network, including resistive random access memories (RRAM)[10], phase change memories (PCM)[11], ferroelectric RAM[12,13] and ferromagnetic memristors[14,15,16]. All exhibit a *nonlinear* dependence of conductance on programming voltage that is specific to the materials and device structure. Unfortunately, the intrinsic nonlinearity of these systems is difficult to customize, and they are not inherently ideal for either linear synaptic functions or programmable nonlinear activation functions. When applied to linear functions, nonlinear nonvolatile devices require additional time to train the accelerators for a particular machine-learning task and cause a loss in learning accuracy[17]. It is also challenging to rely on material and device design to modify the intrinsic response and implement a particular thresholding function between the layers. Such deficiencies force analog data to be converted into digital signals for processing, adding to the latency, energy dissipation, and the physical footprint on-chip required for training and inference operations[18].

We consider an alternative nonvolatile technology. Magnetic domain walls can be manipulated in magnetic wires by applying a current in an adjacent heavy metal. The spin Hall effect in the heavy metal layer creates a spin current that propagates into the magnetic wire, applying spin orbit torque to the domain wall.[19, 20] In the absence of any external magnetic field or pinning of the domain wall, the domain wall velocity increases with the current density up to saturation[21], either expanding or contracting the domain depending on the direction of the applied current. Magnetic domain wall devices have been utilized previously for binary logic[22, 23, 24, 25, 26, 27] and memory[28, 29] applications.

Here, we seek to apply the rich physics of magnetic materials, introducing neuromorphic accelerators based on domain walls that demonstrate both multilevel linear synaptic weight generation and programmable nonlinear thresholding. The devices are implemented using magnetic tunnel junctions (MTJs) connected in parallel on a continuous wire that forms the free layer. The input to the device is current that stores information in the free layer using spin-orbit torque-induced domain wall motion. The output MTJs switch discretely as a function of the input current density. Prototype devices operate with current pulses of 8 ns width and energy consumption of $\leq 16$ pJ. In addition, we demonstrate a proof-of-principle for analog synaptic weight and activation function generation. The design methodology for both discretized and analog devices is described in detail.

**Results**

**Magnetic tunnel junctions characterizations**

To convert the magnetic orientation of a domain back into an electrical signal, we use a perpendicular magnetic tunnel junction (MTJ), consisting of two ferromagnetic CoFeB layers separated by a nonmagnetic MgO insulator. The resistance measured normal to the layers is a function of the relative alignment of the magnetization of the two ferromagnetic layers. In this work, the MTJ thin films employed for the synaptic weight and activation function generators consist of Ta (5 nm)/CoFeB (1 nm)/MgO (2 nm)/CoFeB (1.7 nm)/Ta(0.4 nm)/[Co(0.5 nm)/Pt(1.0 nm)]$_3$/Co(0.5 nm)/Ru (0.9 nm)/Co (0.5 nm)/[Pt (1.0 nm)/Co (0.5 nm)]$_5$/Ru (5 nm), shown schematically in Fig. 1a. The bottom CoFeB layer forms the 'free layer' and supports a programmable domain wall position. The upper part of the stack, including the top CoFeB layer and the synthetic antiferromagnetic (SAF) consisting of two Co/Pt multilayers, forms a reference layer within the MTJ. The inset in Fig. 1a shows a transmission electron microscope image of the crystalline CoFeB and MgO layers after annealing the film stack at 250 °C. The magnetization of all layers is perpendicular to the film plane as shown by vibrating sample magnetometry; see Fig. 1b. The center hysteresis loop confirms the perpendicular anisotropy of the 1-nm-thick CoFeB free layer and the two hysteresis loops on both sides of the center loop represent reversal of the reference CoFeB and the SAF at a field of around 1500 Oe.

Figures 1c and 1d show the scanning electron microscope (SEM) images of a 1 $\mu$m × 3 $\mu$m MTJ and a complete device along with the two terminal resistance measurement setup, respectively. See Methods for details of the device fabrication. Prior to electrical measurements, we apply a large magnetic field of 10 kOe to align the reference layer to a single domain state (↑). Figure 1e shows that the tunnel magnetoresistance (TMR) of a 1 $\mu$m × 3 $\mu$m MTJ is ~ 44% as calculated from a two terminal measurement. From Fig. 1e, it can also be seen that the minor loop of the free layer is not centered at zero magnetic field, which is due to the stray field from the reference layer.

In unpatterned thin films, the free layer of the MTJ stack switches at an applied field of 20 Oe (Fig. 1b). When the MTJ is patterned, however, the switching fields are much larger and dependent on the stray field of the reference layer, as is demonstrated by the large coercive field in the device of Fig. 1e. The stray field is an artifact of our patterned SAF. If the stray field is left uncompensated it can act to trap domain walls, preventing the devices from switching. In Supplementary Note 1 we present alternative designs that effectively eliminate the stray field. However, in this work, we demonstrate the operation of the devices by overcoming the stray field using a compensating external bias field.

Figure 2a is a schematic showing the domain structure in the free layer of a device. Within a range of applied magnetic fields, the regions of the free layer in the wire without reference layers align parallel to the field while the regions with reference layers remain antiparallel to it – creating domain walls inside the CoFeB free layer at the edges of the reference layers. The domain structures within the CoFeB free layer are examined by probing anomalous Hall resistance within and outside the MTJ region, as shown in Fig. 2c and 2d, respectively using the electrical set-up shown in Fig. 2b. A large difference in the corresponding switching fields for the two regions is evident, consistent with the domain pictures described above.

To characterize current-induced domain wall motion inside the free layer of a single MTJ, we apply a bias field perpendicular to the sample, then we apply a series of current pulses of 20 ns width with increasing amplitude and measure the TMR after each pulse. The measurement setup is shown in Fig. 3a. The bias field serves two purposes – firstly, to nucleate domain walls at the edges of the tunnel junction and secondly, to counteract the stray field from the reference layers.

Figure 3b shows the complete switching of a 200 nm × 400 nm MTJ from parallel to antiparallel state at different bias fields. The pinning field from the reference layer is maximum at the edges of the reference layer[30]. A combination of bias field and spin torque from the current pulse translates the domain wall under the MTJ, switching the state of the MTJ. The current density required to switch the MTJ increases with a decrease in the amplitude of the bias field; see Fig. 3b. The switching current densities extracted from experiments on MTJs with different widths are consistent with a uniform threshold for domain wall depinning in a CoFeB wire[20]; see Supplementary Note 2. Because of the existence of magnetic domain walls at opposite edges of each MTJ, a current in either direction can cause domain wall propagation and change the state of the device. Further characterization of this device is shown in Supplementary Note 3. After the output state is reached, the system is reset by the application of a larger bias field. The free layer could be reset by applying a vertical current through the MTJs or by propagating a domain wall from one end of the wire to change the magnetization of the free layer except under the MTJs.

**Discretized synaptic weight generation**

We next design and characterize a synaptic function generator with a linear relationship between the input current and output resistance; see Supplementary Note 4. We pattern a series of discrete MTJs each with dimensions 200 nm × 400 nm on a CoFeB free layer. The linear function is implemented by patterning the free layer (bottom CoFeB) and the underlying heavy metal (Ta) with linearly varying width, which changes the current density linearly along the device. Figures 3c and 3d-e show the schematic and SEM images of the patterned structure, respectively. All MTJs are connected in parallel. With a suitable bias field, each of the MTJs supports a discrete magnetic domain as shown in Fig. 3c.

TMR measurements as a function of magnetic field yield steps in resistance *vs*. field, which correspond to the switching of nine MTJs from parallel to antiparallel states (Supplementary Note 3). Each MTJ switches at a slightly different magnetic field due to fabrication-induced inhomogeneities. To measure the current-driven switching, we apply a series of 8-ns current pulses along the CoFeB/Ta wire with increasing amplitudes and observe the increase in magnetoresistance across the MTJs. Figure 3f shows the resistance of the parallel MTJs as a function of both positive and negative current in the CoFeB/Ta wire. The eight different resistance values correspond to consecutive switching of seven MTJs. The systematic change in resistance and the corresponding domain wall positions are shown in Supplementary Fig. 7. We did not observe the switching of the remaining two MTJs on the widest part of the wire. The current required to switch them is higher than the breakdown current of the narrowest part of the wire. The dotted lines show that the resistance of the synaptic device changes linearly with the applied current – an ideal relationship for direct mapping of the synaptic weights in the algorithms to the conductance of the device.

**Discretized activation function generation**

To implement a nonlinear thresholding function, we again pattern a series of MTJs on a CoFeB/Ta wire. In addition to varying the width of the CoFeB/Ta wire, we vary the MTJ areas, with large devices in the center of the wire, corresponding to the steepest portion of the activation characteristic. The thresholding device in this particular example is chosen to approximate a sigmoid function, a traditional nonlinear function for deep neural networks[2]. The CoFeB/Ta wire is patterned with a linear increase in width along its length to initiate the switching of each MTJ at regularly-spaced input currents. Like the synaptic weight generator, all the MTJs are connected in parallel. The detailed design of the activation function generator is discussed in the Supplementary Note 5.

Figures 4a and 4b show a schematic and SEM images of the fabricated activation function generator, respectively. We first characterize the nonlinear MTJ device by applying an out-of-plane magnetic field and observing the field induced switching of the nine MTJs in the resistance *vs*. field plot; see Supplementary Note 6. The MTJ switching from parallel to antiparallel and from antiparallel to parallel occurs asymmetrically with applied field due to magnetostatic interaction between the free layer and reference layer. We also observe stochasticity in the switching fields in repeated magnetoresistance measurements. Next, we use the experimental setup of Fig. 3a to measure the electrical characteristics, with input current pulses ramped from zero to 620 $\mu$A. The bias field is chosen to be 200 Oe allowing us to observe switching of the maximum number (seven) of MTJs. The remaining two devices switch at lower fields. Averaged over 75 cycles to reduce thermal noise, the resistance of the parallel MTJs when switching with increasing current is shown in Fig. 4c. The experimental data agrees well with the designed response of the device with seven MTJs, and follows the target shifted sigmoid activation function.

**Analog implementation of activation function**

The approach used above stores information digitally in multiple, discrete magnetic domains positioned along a magnetic nanowire. The magnetization of the free layer under each MTJ can be reversed by spin orbit torque-driven domain wall motion to change the device resistance stepwise according to the desired function. The projected information density of these devices is expected to approach that of MRAM. It is possible, however, to also consider an analog implementation of these neuromorphic devices with higher densities. In an analog device, the position of a domain wall in an MTJ wire is a quasi-continuous variable, with spatial resolution fundamentally limited by the correlation length of the patterning process employed during fabrication[31]. As characterized in Supplementary Note 6, the domain wall can be moved in either direction by the application of a current. We also develop an analytical model for designing an activation function utilizing spin orbit torque-induced domain wall motion (Supplementary Note 8[18]). By modifying the width of the heavy metal wire carrying the current, our micromagnetic simulation confirms that domain wall travelling distances can be a nonlinear function of the input current.

To enable quasi-analog operation, we use a Pt (3 nm)/CoTb (2 nm)/SiNx (3 nm) device with non-uniform width without a magnetic reference layer such that the domain wall can reside at any position along the CoTb wire. Magnetooptic Kerr effect measurements are used to characterize domain wall motion. We pattern the Pt wire in a shape determined from the analytical function (Supplementary Note 8) and keep the CoTb wire width constant; see Fig. 5a. The details of the fabrication steps are discussed in Methods. Using the magneto-optic Kerr effect and electrical setup described in Methods, we measure the domain wall motion after applying current pulses of width 300 ns; see Fig. 5b. For each measurement, the domain walls are initialized at A as shown in the inset of Fig. 5b. We observe that the domain wall travelling distances successfully follow the sigmoid function design for the activation function generator. To implement an analog function generator with electrical output, a continuous MTJ reference layer can be designed on top of the magnetic wire to enable electrical detection of domain wall positions[14, 32].

**Discussion**

The linear and nonlinear behavior exhibited by the devices in Figs. 3, 4 and 5 demonstrates that three terminal MTJs can be customized for multiple roles within a neuromorphic accelerator. In the discretized devices, optimization of the functional behavior could also occur post-fabrication. In the same way that information is stored in magnetic random access memories (MRAMs), it is possible to drive current through individual MTJs to selectively program particular steps in the desired magnetoresistance curve by initializing each magnetic domain prior to applying the input current[33, 34, 35]. Individual tunnel junctions[34, 35] without continuous free layers and domain walls can also implement the linear and the nonlinear device characteristics if the width of the heavy-metal-wire and areas of tunnel junctions are engineered.

A key consideration for application to neuromorphic systems is power consumption. Considering the resistive loss in the bottom CoFeB/Ta wire with resistance of 5.2 k$\Omega$, the power required to switch the individual MTJs in the prototype linear and nonlinear devices varies from 0.15 mW to 2.0 mW. With an 8 ns current pulse, the energy consumption ranges from 1 pJ to 16 pJ (see Supplementary Note 9 for the detail analysis of switching time of the MTJs). For a scaled 20-nm-wide MTJ, the width of the bottom CoFeB wire can be as narrow as 25 nm with a corresponding increase in the cross sectional area. In 25-nm-wide wires, however, the current density for domain wall motion can be 2-4 times higher than that required in our prototype devices[36]. Given the same 8 ns current pulses employed here, the expected energy consumption per synaptic event is therefore ~18-36 fJ. At such low switching energies it is conceivable that the retention time will be less than the conventional 10 year MRAM benchmark. Thus, assuming reduced retention times are feasible in neuromorphic applications, we expect that the energy consumption in a 20 nm wide MTJ will be lower than other nonvolatile devices and within one order of magnitude of biological synapses[37].

A final advantage is that MTJ-based devices are the basis of MRAMs and are compatible with traditional silicon processing. Unlike the requirements for logic devices, the resistance of synaptic devices must be sufficiently large to ensure that each device in a cross bar architecture contributes only a small amount of current. For application into cross-point array architectures, the resistances of the synaptic weight generators are required to be in the range of mega ohms[38]. This can be achieved in the proposed design by increasing the thickness of the MgO tunnel barrier if the size of the MTJs are chosen the same as shown in Fig. 3d. However, the thickness of the MgO barrier should be 1.6 nm – 1.8 nm for a scaled MTJ of 20 nm × 10 nm, considering the resistance-area product[39]. In cases, where MTJs are utilized to drive complex circuits, much lower thickness of the MgO layer will ensure the proper scaling of the device[40].

To summarize, we have demonstrated a linear synaptic weight generator and a nonlinear activation function generator using domain wall motion under a series of magnetic tunnel junctions. Our prototype devices operate with 8 ns current pulses with energy consumption on the order of 1-16 pJ. The energy required by scaled devices is predicted to be very close to that of biological synapses. Furthermore, the proposed devices are compatible with silicon foundry processes. The versatility of magnetic domain wall technology for both computation and memory with very low latency and energy-consumption suggests that it is an excellent candidate for use in neuromorphic accelerators.

## Methods

### Sample preparation

The thin metal films for the magnetic tunnel junction structures are grown by DC magnetron sputtering and the tunnel barrier, MgO is deposited by radio frequency sputtering at 2 mTorr on a thermally oxidized Si substrate. The base pressure of the chamber is $8\times10^{-9}$ Torr. The thin film stack is annealed in a $N_2$ environment at 250 °C for 15 min on a hotplate. Fabrication of the devices is completed using three electron-beam lithography and two $Ar^+$ ion-milling steps. In the first lithography step, we pattern the free layer (bottom wires) and ion-mill the complete stack. In the next lithography step, we pattern the tunnel junctions and ion-mill up to the MgO barrier using an end-point detector. After this step, we evaporate $SiO_2$ to ensure the electrical separation of the top and the bottom contacts of the tunnel junctions. In the last lithography step, the contacts are patterned. We evaporate Ti (10 nm)/Au (100 nm) for the contacts of the wires and the tunnel junctions and lift-off the metals afterwards.

The CoTb samples are deposited using DC magnetron co-sputtering of Co and Tb. The films are capped with 3 nm of $SiN_x$ insulating layer using RF sputtering. The samples for magneto optic Kerr effect measurement are patterned using a bilayer resist stack of Hidogensilexoquizen (HSQ) and PMMA in two e-beam lithography steps. In the first e-beam lithography step, we pattern the bottom Pt wire. The films are then ion-milled with $Ar^+$ ions. We remove the resists stack and spin the same resists stack again for the second e-beam lithography step. In this step, the $CoTb/SiN_x$ wires with constant widths are patterned and ion-milled afterwards. Finally, we remove the resists stack in hot NMP. We ensure that the processing temperature of the CoTb samples remains < 100 °C.

**Electrical measurement**

The anomalous Hall resistance of the wires is measured using four contacts and the magnetoresistance of the tunnel junctions is measured using two contacts. A Keithley 2400 source meter is utilized for these quasi DC measurements. The magnetic tunnel junctions are measured on a custom built probe station equipped with an out-of-plane electromagnet. To measure the current induced domain wall motion, we use radiofrequency probe for applying the nano-second current pulses in CoFeB wires using either an Agilent 8130A or an Agilent 8114A pulse generator and utilize a DC probe to measure the tunnel magnetoresistance using a Keithley 2400. The current applied to measure the tunnel junction was 100 nA.

For magneto-optic Kerr effect measurements, the fabricated devices are bonded onto a radio frequency chip carrier with cut-off frequency of 5 GHz. An Agilent 8114A pulse generator is employed to provide the nanosecond current pulses. The images are captured before and after the applied pulses with a custom-built set up equipped with in-plane and out-of-plane magnets.

**Acknowledgements**

This project was supported by the National Science Foundation award 1639921, and the Nanoelectronics Research Corporation, a subsidiary of the Semiconductor Research Corporation, through Memory, Logic, and Logic in Memory Using Three Terminal Magnetic Tunnel Junctions, an SRCNRI Nanoelectronics Research Initiative under Research Task ID 2700.002. This work also used the MIT MRL (formerly CMSE) Shared Experimental Facilities supported by DMR 1419807 and the shared facilities at MIT NanoStructures Laboratory and Microsystems Technology Laboratories.

**Author Contributions**

M.A.B. conceived the project. S.A.S. designed & fabricated the devices, performed electronic and optical transport measurements and analyzed the data. S.D. modeled the single domain wall devices. A.T. prepared the TEM sample. L.L., C.A.R., and M.A.B. supervised the project and provided scientific support. S.A.S., and M.A.B. co-wrote the manuscript with input from all co-authors.

**Competing interests**

The authors declare no competing interests.

**Data availability**

The data that support the plots within this paper and other findings of this study are available from the corresponding authors upon reasonable request.

**Corresponding authors**

Correspondence to Saima A Siddiqui or Marc A Baldo.

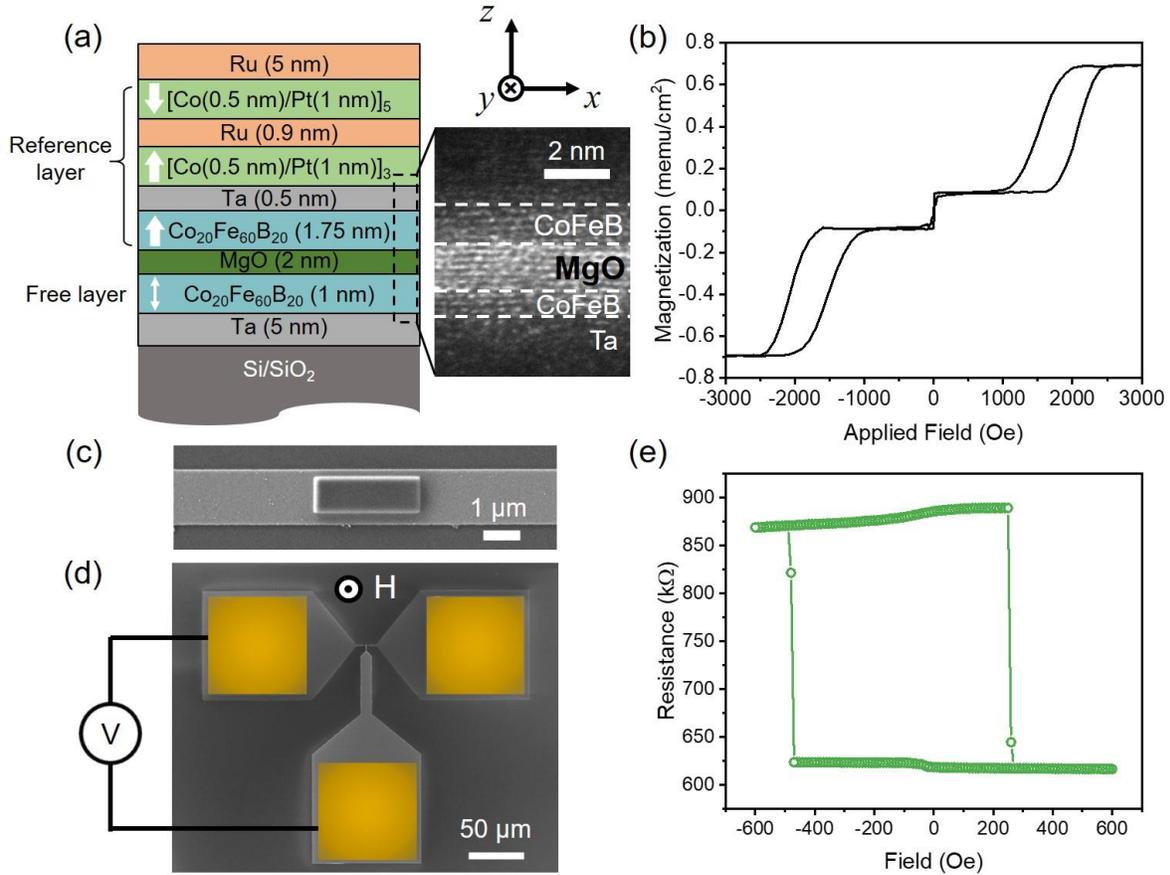

Figure 1| **Tunnel junction for measuring magnetoresistance. a,** Thin film structure for fabricating magnetic tunnel junctions. The 1-nm thick bottom $Co_{20}Fe_{60}B_{20}$ layer serves as a free layer for the tunnel junctions. The top 1.75-nm-thick $Co_{20}Fe_{60}B_{20}$ layer is coupled parallel to the bottom Co/Pt multilayer through a 0.5 nm thick Ta layer and the top and bottom Co/Pt multilayers are antiferromagnetically coupled through the 0.9 nm thick Ru layer. The transmission electron microscope image shows structures of thin films inside the dotted box. **b,** Out-of-plane hysteresis loop of the annealed thin film structure. The center loop shows the switching of the free layer and the two loops on the two sides of the center loop show the switching of the synthetic antiferromagnet and the CoFeB reference layer at ~1500 Oe. **c, d,** Scanning electron microscope images of a 1 μm × 3 μm MTJ without (**c**) and with (**d**) electrical contacts, including the electrical

schematic for measuring the magnetoresistance of the fabricated tunnel junctions. **e**, Magnetoresistance of the device in **d**. Magnetic field is applied out of the plane of the junction.

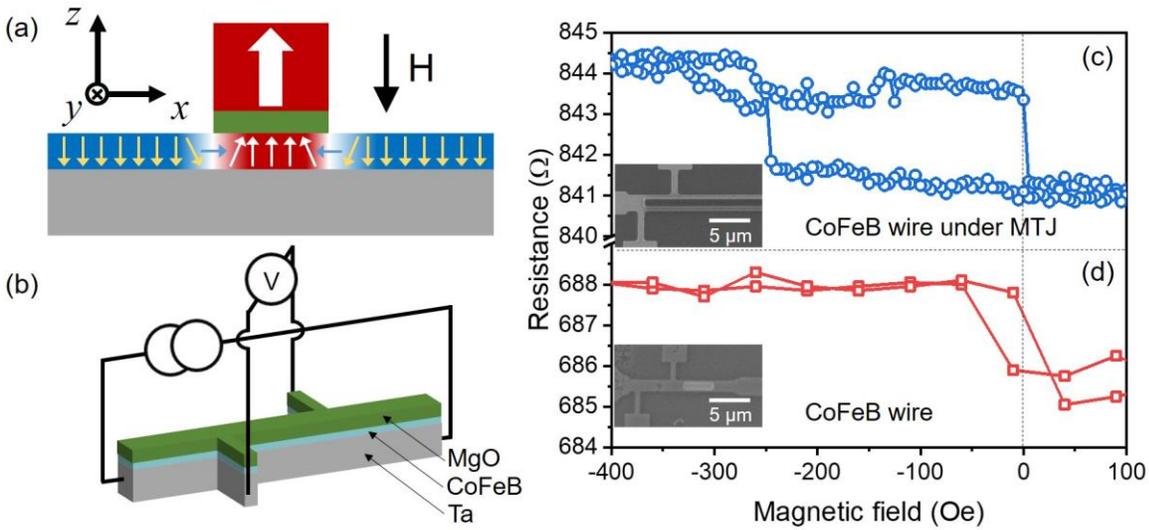

Figure 2| **Magnetic domains in the free layer of a tunnel junction. a,** Schematic of the fabricated tunnel junction after initializing the soft layer with a large field in the +*z* direction and then applying a small field along –*z* direction. The blue region of the wire shows the down domain (marked by yellow arrows) and the red region of the wire shows the up domain (marked by white arrows). The region of the wire under the tunnel junction remains parallel to the reference layer orientation. The rest of the wire is magnetically softer and switches with a small magnetic field. Thus, domain walls are nucleated at the sides of the tunnel junctions. **b** Schematic of the anomalous Hall resistance measurement set-up for probing the coercive fields of the free layers with and without reference layers. A magnetic field is swept in ±*z* and the voltage is measured on the two terminals of the device at every field with a constant current on the other two terminals. **c, d,** Magnetic switching of regions of the CoFeB wire with and without tunnel junctions, probed via the Hall resistance of the free layer. The contacts for the Hall voltage are located within and outside the MTJ region in **c** and **d**, respectively. The hysteresis loop of the CoFeB under the tunnel junction is offset from zero due to the stray field from the CoFeB reference layer.

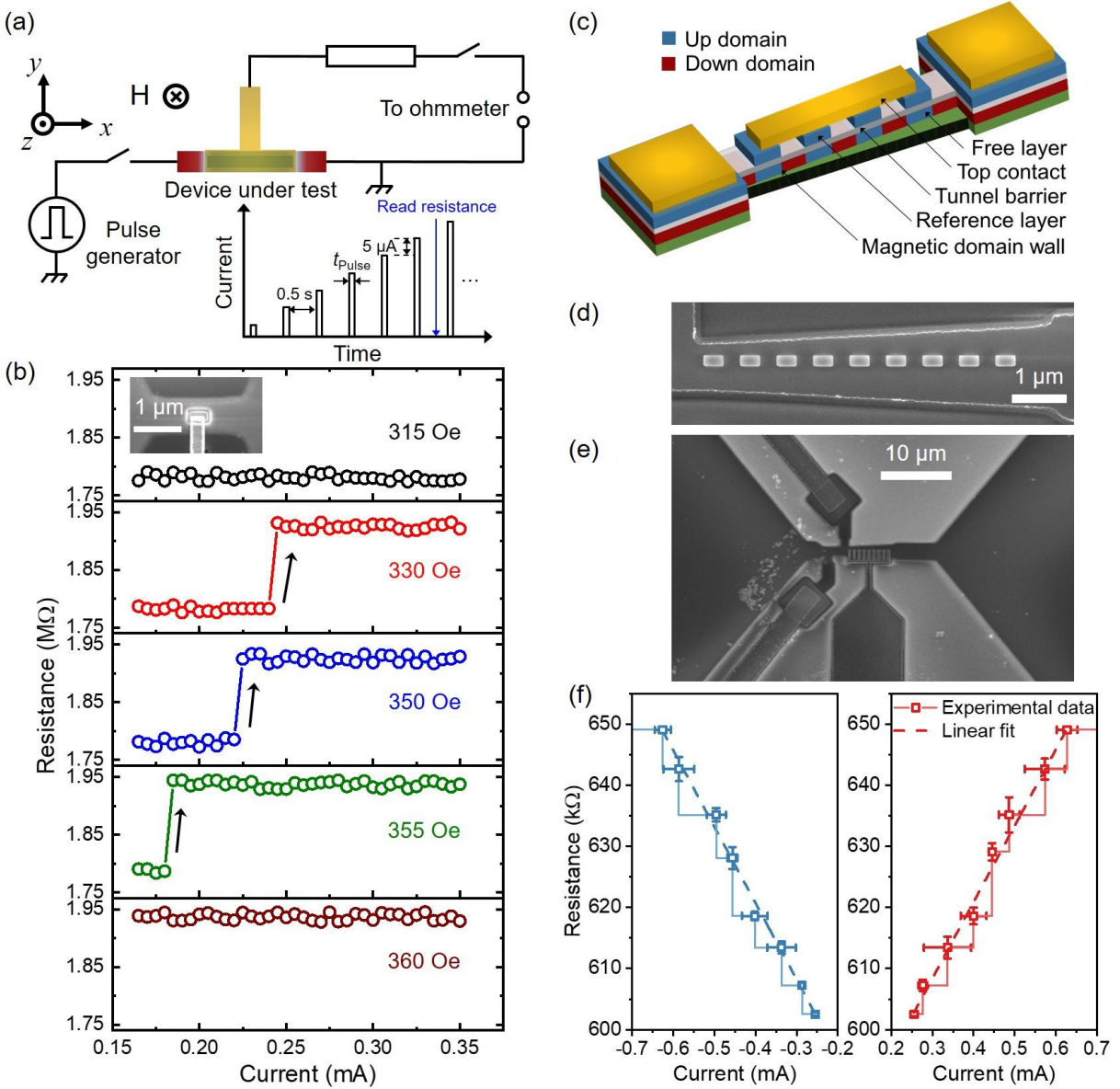

Figure 3 **Synaptic weight generator using magnetic tunnel junctions. a,** Schematic top view of the magneto-electrical set-up for measuring the magnetoresistance of the magnetic tunnel junctions with magnetic domain walls. A bias field lower than the switching field of the free layer is applied out of the plane of the junction to reduce the pinning of the domain walls. Inset shows the input pulse trains from the pulse generator. The resistance of the tunnel junction device is measured after each pulse. **b,** Magnetoresistance of a 200 nm × 400 nm MTJ with $t_{Pulse}$ = 20 ns at different bias fields. The arrows show the direction of change in magnetoresistance. The switching current is reduced at larger bias fields. **c,** Schematic of a series of tunnel junctions device for multi-level discretized synaptic weight generation. The boundaries between the blue (up domain) and the red (down domain) regions in the free layer represent the domain walls. **d, e,** Scanning electron microscope images of nine 200 nm × 400 nm MTJs without (d) and with (e) electrical contacts. The tunnel junctions are connected in parallel to each other. **f,** Switching of the parallelly connected MTJs in **e** with a bias field of – 480 Oe. The eight steps correspond to the switching of seven MTJs with $t_{Pulse}$ = 8 ns. The linear fits of the magnetoresistance confirm the linear weight generation using MTJs. The error bars of resistances are obtained by repeating the measurements several times in a single device.

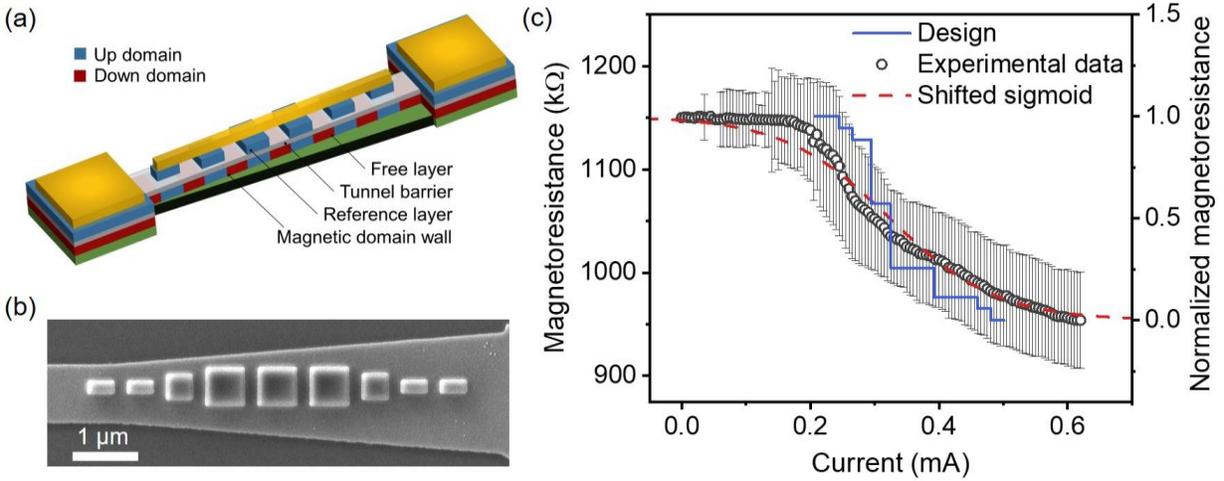

Figure 4| **Activation function generator with multiple domain walls and magnetic tunnel junctions. a,** Schematic of a series of tunnel junctions for discretized activation function generation. The boundaries between the blue (up domain) and the red (down domain) regions in the free layer represent the domain walls. **b,** Scanning electron microscope images of nine magnetic tunnel junctions. The width of the free layer is varying linearly along its length. To implement the nonlinearilty, areas of the MTJs are varied along the length of the free layer since the resistance modulation is proportional to the size of the MTJs (Supplementary Note 1). **c,** Switching of in-parallel magnetic tunnel junctions in **b** with a bias field of 200 Oe. The magnetoresistances are averaged from 75 measurements with 8 ns current pulses and the error bars show the standard deviation of magnetoresistance. The solid blue line shows the response of the designed activation generator when a 600 nm × 600 nm (5th MTJ from the left) and a 400 nm × 400 nm (7th MTJ from the left) MTJs are already switched by the bias field of 200 Oe which fits best with the observed data. For the design, the switching current density is assumed to be $0.7 \times 10^{11}$ J/m$^2$.

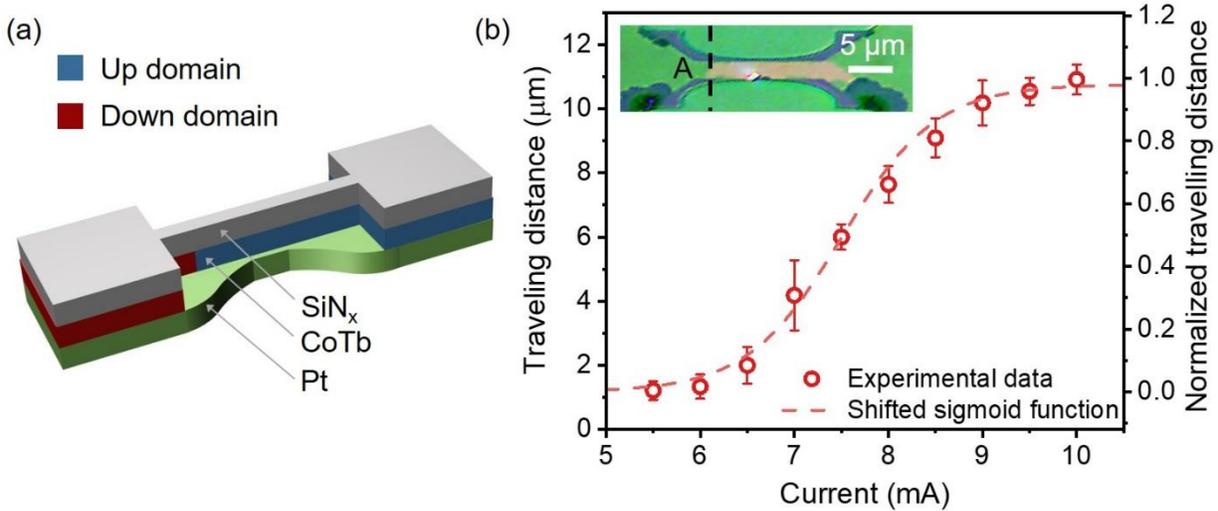

**Figure 5| Analog activation function generator with single magnetic domain wall. a,** Schematic of an activation function generator with a single domain wall in a CoTb wire. The bottom Pt layer is patterned so that its width varies along the length of the wire according to the analytical model obtained from a shifted sigmoid function[18]; see Supplementary Note 7. The width of the CoTb wire is constant. **b,** Magnetic domain wall travelling distances in a CoTb wire as shown in the inset with 300 ns current pulses. Inset shows the magnetoptic Kerr microscope image of a CoTb wire with two domain walls. The black dotted line marked by **A** in the inset shows the initial position of the domain wall for all currents.